\documentclass[11pt,letterpaper,showkeys]{article}
\usepackage{graphicx}
\usepackage{amssymb,amsmath,mathrsfs}
\usepackage{amsthm,enumerate,verbatim}
\newtheorem{theorem}{Theorem}[section]

\newtheorem{proposition}[theorem]{Proposition}
\newtheorem{remark}[theorem]{Remark}

\newcommand{\beq}{\begin{equation}}
\newcommand{\eeq}{\end{equation}}
\newcommand{\beqa}{\begin{eqnarray}}
\newcommand{\eeqa}{\end{eqnarray}}
\newcommand{\beqas}{\begin{eqnarray*}}
\newcommand{\eeqas}{\end{eqnarray*}}
\newcommand{\ba}{\begin{array}}
\newcommand{\ea}{\end{array}}
\newcommand{\bi}{\begin{itemize}}
\newcommand{\ei}{\end{itemize}}

\def\argmin{\mathop{\rm argmin}}
\def\Argmin{\mathop{\rm Argmin}}

\def\QED{\ifhmode\unskip\nobreak\fi\ifmmode\ifinner\else\hskip5pt\fi\fi
  \hbox{\hskip5pt\vrule width5pt height5pt depth1.5pt\hskip1pt}}
\title{An Alternating Direction Method for Finding Dantzig Selectors}

\date{November 19, 2010}

\author{
    Zhaosong Lu%
    \thanks{
    Department of Mathematics, Simon Fraser University, Burnaby, BC,
    V5A 1S6, Canada. (email: {\tt zhaosong@sfu.ca}). This author was
    supported in part by NSERC Discovery Grant.}
    \and
    Ting Kei Pong
    \thanks{Department of Mathematics, University of Washington,
    Seattle, Washington 98195, U.S.A. (email: {\tt tkpong@math.washington.edu}).}
    \and
    Yong Zhang
    \thanks{Department of Mathematics,
    Simon Fraser University, Burnaby, BC, V5A 1S6,
    Canada. (email: {\tt yza30@sfu.ca}).}
}

\begin{document}

\maketitle

\begin{abstract}

In this paper, we study the alternating direction method for
finding the Dantzig selectors, which are first introduced in \cite{CaTa05}.
In particular, at each iteration we apply the nonmonotone gradient method
proposed in \cite{LuZh09} to approximately solve one subproblem of this method.
We compare our approach with a first-order method proposed in \cite{BCG10}.
The computational results show that our approach usually outperforms that
method in terms of CPU time while producing solutions of comparable quality.

\vskip14pt


\vskip14pt

%\noindent
%{\bf AMS 2000 subject classification:}

\end{abstract}

\section{Introduction}\label{sec:intro}

Consider the standard linear regression model:
\begin{equation}\label{linear}
y=X\beta+\epsilon,
\end{equation}
where $y\in \Re^n$ is a vector of responses, $X\in \Re^{n\times p}$
is a design matrix, $\beta\in \Re^p$ is an unknown regression vector
and $\epsilon$ is a vector of random noises. One widely studied
problem for this model is that of variable selection, that is, how
to determine the support of $\beta$ (i.e., the indices of the
nonzero entries of $\beta$). When $p\ll n$, this problem can be
tackled by many classical approaches. In recent years, however, the
situations where $p\gg n$ have become increasingly common in many
applications such as signal processing and gene expression studies.
Thus, efforts have been directed at developing new variable
selection methods that work for large values of $p$. A few examples
of such methods include the lasso \cite{Tib96}, the elastic net
\cite{ZoH05}, and the more recent Dantzig selector \cite{CaTa05}.

A Dantzig selector for \eqref{linear} is a solution of the following
optimization problem:
\begin{equation}\label{P_orig}
  \begin{array}{rl}
    v=\displaystyle\min_{\beta}& \|\beta\|_1\\
    {\rm s.t.}& \|D^{-1}X^T(X\beta-y)\|_{\infty}\le \delta,
  \end{array}
\end{equation}
where $\delta>0$ and $D$ is the diagonal matrix whose diagonal
entries are the norm of the columns of $X$. The Dantzig selector was
first proposed in \cite{CaTa05} and justified on detailed
statistical grounds. In particular, it was shown that, this
estimator achieves a loss within a logarithmic factor of the ideal
mean squared error, i.e., the error one would achieve if one knows
the support of $\beta$ and the coordinates  of $\beta$ that exceed
the noise level. For more discussion of the importance of Dantzig
selector and its relationship with other estimators like lasso, we
refer the readers to
\cite{Bic07,CaL07,CaTa07b,EHT07,FrSa07,MRY07,Rit07,JRL09}.

Despite the importance of the Dantzig selector and its many
connections with other estimators, there are very few existing
algorithms for solving \eqref{P_orig}. One natural way of solving
\eqref{P_orig} is to recast it as a linear programming (LP) problem and
solve it using LP techniques. This approach is adopted in the package
$\ell_1$-magic \cite{CaRo05}, which solves the resulting LP problem via
a primal-dual interior-point (IP) method. However, the IP methods are
typically not efficient for large-scale problems as they require solving
dense Newton systems for each iteration. Another approach of solving
\eqref{P_orig} uses homotopy methods to compute the entire solution
path of the Dantzig selector (see, for example, \cite{Rom08,JRL09}).
Nevertheless, as discussed in \cite[Section~1.2]{BCG10}, these methods are also
unable to deal with large-scale problems. Recently, first-order methods
are proposed for \eqref{P_orig} in \cite{Lu09,BCG10}, which are capable
of solving large-scale problems. In \cite{Lu09}, problem~\eqref{P_orig}
and its dual are recast into a smooth convex programming problem and an
optimal first-order method proposed in \cite{AuT06} is then applied to
solve the resulting problem.
%However, the dimension of the latter problem is at least doubled after the
% reformulation, which can be a disadvantage when solving large-scale
% problems.
In \cite{BCG10}, problem~\eqref{P_orig} is recast as a linear cone
programming problem. The optimal first-order methods (see, for
example, \cite{Nes83,AuT06,Nes07,Tse08,LLM09}) are then applied to
solve a smooth approximation to the dual of the latter problem.
% Each iteration of these methods is
% typically cheap. However, the total number of iterations needed to
% solve the smooth approximation up to a certain accuracy is typically
% inversely proportional to the smoothing parameter \cite{Nes05},
% which has to be small so that the dual problem is well approximated
% by the smooth approximation. Thus, these methods can be slow in
% practice.

In this paper, we consider an alternative approach, namely, the alternating
direction method (ADM), for solving \eqref{P_orig}.  The ADM and its many
variants have recently been widely used to solve large-scale problems in compressed
sensing, image processing and statistics (see, for example, \cite{ABF09,EZC09,YaZ09,Yua09,
YZY10}). In general, the ADM can be applied
to solve problems of the following form:
\begin{align}\label{form}
  \begin{array}{rl}
    \displaystyle\min_{x,y}& f(x) + g(y)\\
    {\rm s.t.}& Ax + By = b,\\
    & x\in C_1, y\in C_2,
  \end{array}
\end{align}
where $f$ and $g$ are convex functions, $A$ and $B$ are matrices, $b$ is a
vector, and $C_1$ and $C_2$ are closed convex sets. Each iteration of the
ADM involves solving two subproblems successively and then updating a multiplier,
and the method converges to an optimal solution of \eqref{form} under some  mild
assumptions (see, for example, \cite{BeTs89,EcB92}). In this paper,
we show that \eqref{P_orig} can be rewritten in the form of
\eqref{form}, and hence the ADM can be suitably applied. Moreover, we show that
one of the ADM subproblems has a simple closed form solution, while another one can be
efficiently and approximately solved by a nonmonotone gradient method proposed
recently in \cite{LuZh09}. We also discuss convergence of this ADM. Finally, we compare our
method for solving \eqref{P_orig} with a first-order method proposed in \cite{BCG10}
on large-scale simulated problems. The computational results show that our approach
usually outperforms that method in terms of CPU time while producing solutions of
comparable quality.

The rest of the paper is organized as follows. In Subsection~\ref{sec:notation},
we define notations used in this paper. In Section~\ref{sec:dant}, we study the
alternating direction method for solving problem~\eqref{P_orig} and address its
convergence. Finally, we conduct numerical experiments to compare our method with
a first-order method proposed in \cite{BCG10} in Section~\ref{sec:numerics}.
% Finally, we present some concluding remarks in Section~\ref{sec:conclude}.

\subsection{Notations}\label{sec:notation}

In this paper, $\Re^n$ denotes the $n$-dimensional Euclidean space
and $\Re^{m\times n}$ denotes the set of all $m \times n$ matrices
with real entries. For a vector $x\in\Re^n$, $\|x\|_1$, $\|x\|_2$
and $\|x\|_\infty$ denote the $1$-norm, $2$-norm and $\infty$-norm
of $x$, respectively. For any vector $x$ in $\Re^n$, $|x|$ is the
vector whose $i$th entry is $|x_i|$, while ${\rm sgn}(x)$ is the
vector whose $i$th entry is $1$ if $x_i>0$ and $-1$ otherwise. Given
two vectors $x$ and $y$ in $\Re^n$, $x\circ y$ denotes the Hadamard
(entry-wise) product of $x$ and $y$, $\max\{x,y\}$ denotes the
vector whose $i$th entry is $\max\{x_i,y_i\}$. The letter $e$
denotes the vector of all ones, whose dimension should be clear
from the context. Finally, given a scalar $a$, $[a]_+$ denotes
the positive part of $a$, that is, $[a]_+ = \max\{0,a\}$.

\section{Alternating direction method}\label{sec:dant}

In this section, we study the ADM for solving \eqref{P_orig} and
discuss its convergence and implementation details.

In order to apply the ADM, we first rewrite \eqref{P_orig} in the
form of \eqref{form}. To this end, we introduce a new variable $z$
and rewrite \eqref{P_orig} as follows:
\begin{equation}\label{P_deriv}
  \begin{array}{rl}
    \min\limits_{\beta}& \|\beta\|_1\\
    {\rm s.t.}& X^T(X\beta-y)-z=0,\\
    & \|D^{-1}z\|_{\infty}\le \delta.
  \end{array}
\end{equation}
Then it is easy to see that \eqref{P_deriv} is in the form of
\eqref{form} with $A=X^TX$, $B=-I$, $c=X^Ty$, $C_1=\Re^p$,
$C_2=\{z:\;\|D^{-1}z\|_{\infty}\le \delta\}$, $f=\|\cdot\|_1$ and
$g=0$. Next, in order to describe the ADM iterations, we introduce
the following augmented Lagrangian function for problem \eqref{P_deriv}:
\begin{equation*}
  L_\mu(z,\beta,\lambda) =
  \|\beta\|_1+\lambda^T(X^TX\beta-X^Ty-z)+\frac{\mu}{2}\|X^TX\beta-X^Ty-z\|_2^2
\end{equation*}
for some $\mu>0$.
Each iteration of the ADM involves alternate minimization of $L_\mu$
with respect to $z$ and $\beta$, followed by an update of $\lambda$.
The standard ADM for problem \eqref{P_deriv} (or, equivalently, \eqref{P_orig})
is described as follows:\\

\noindent {\bf Alternating direction method:}
\begin{itemize}
  \item[1.] {\bf Start:} Let $\beta^0,\lambda^0\in \Re^p$ and $\mu>0$ be given.
  \item[2.] {\bf For} $k=0, 1, \ldots$
  \begin{equation}\label{subproblem}
  \begin{cases}
    z^{k+1} = \displaystyle \argmin_{\|D^{-1}z\|_\infty\le \delta} L_\mu(z,\beta^k,\lambda^k),\\
    \beta^{k+1} \in\  \displaystyle\Argmin_{\beta}\ L_\mu(z^{k+1},\beta,\lambda^k),\\
    \lambda^{k+1} =\lambda^k+\mu(X^TX\beta^{k+1}-X^Ty-z^{k+1}).
  \end{cases}
  \end{equation}
  {\bf End} (for)
\end{itemize}

Before discussing the convergence of the above method, we first
derive the dual problem of \eqref{P_deriv} (or, equivalently,
\eqref{P_orig}). Note that
\begin{align*}
  v &\ =\ \min_{z,\beta}\{\|\beta\|_1:\;X^T(X\beta-y)-z=0, \|D^{-1}z\|_{\infty}\le \delta\}\\
  &\ =\ \min_{z,\beta}\max_{\lambda}\{\|\beta\|_1+\lambda^T(X^TX\beta-X^Ty-z):\;\|D^{-1}z\|_{\infty}\le
  \delta\}\\
  &\ =\ \max_{\lambda}\min_{z,\beta}\{\|\beta\|_1+\lambda^T(X^TX\beta-X^Ty-z):\;\|D^{-1}z\|_{\infty}\le
  \delta\}\\
  &\ =\ \max_{\lambda}\{-y^TX\lambda-\delta\|D\lambda\|_1:\;\|X^TX\lambda\|_{\infty}\le
  1\},
\end{align*}
where the third equality holds by strong duality. Thus, the dual
problem of \eqref{P_deriv} is given by
\begin{equation}\label{D_orig}
  \begin{array}{rl}
    \max\limits_\lambda& d(\lambda):=-y^TX\lambda-\delta\|D\lambda\|_1\\
    {\rm s.t.}& \|X^TX\lambda\|_{\infty}\le 1.
  \end{array}
\end{equation}
Now we are ready to state a convergence result for the ADM, whose
proof can be found in \cite{BeTs89}.
\begin{proposition}\label{convergence}
  Suppose that the solution set of \eqref{P_orig} is nonempty and $\mu>0$. Let
  $\{(z^k,\beta^k,\lambda^k)\}$ be a sequence generated from the above
  alternating direction method. Then $\{(z^k,\lambda^k)\}$ is convergent.
  Furthermore, the limit of $\{\lambda^k\}$ solves
  \eqref{D_orig}, and any accumulation point of $\{\beta^k\}$ solves
  \eqref{P_orig}.
\end{proposition}

% We next discuss how to solve the two subproblems in the ADM as detailed in \eqref{subproblem}. First,
It is easy to observe that the first subproblem in \eqref{subproblem} has a closed form solution,
which is given by:
\[z^{k+1}=\displaystyle\argmin_{\|D^{-1}z\|_\infty\le \delta}\left\|z-\left(X^TX\beta^k-X^Ty+\frac{\lambda^k}{\mu}\right)\right\|_2^2
=\min\left\{\max\left\{X^TX\beta^k-X^Ty+\frac{\lambda^k}{\mu},-\delta
d\right\},\delta d\right\},\] where $d$ is the vector consisting of the diagonal entries of $D$. However,
the second subproblem does not in general have a closed form solution. In practice we can choose $\beta^{k+1}$
to be a suitable approximate solution instead.  Our next proposition states that the resulting ADM still converges to optimal
solutions. The proof follows essentially the same arguments as \cite[Theorem~8]{EcB92} and is thus omitted.

\begin{proposition}
  Suppose that the solution set of \eqref{P_orig} is nonempty and $\mu>0$. Let $\{\nu_k\}$ be a sequence of nonnegative numbers
  with $\sum \nu_k<\infty$. Let
  $\{(z^k,\lambda^k)\}$ be generated as in \eqref{subproblem} while $\{\beta^k\}$ is chosen to satisfy:
  \[\inf\left\{\|\beta^k-\beta\|_2:\;\beta\in{\rm Argmin}\, L_\mu(z^{k},\beta,\lambda^{k-1}))\right\}\le \nu_k\]
  for all $k$. Then $\{(z^k,\lambda^k)\}$ is convergent. Furthermore, the limit of $\{\lambda^k\}$ solves
  \eqref{D_orig}, and any accumulation point of $\{\beta^k\}$ solves
  \eqref{P_orig}.
\end{proposition}

Before ending this section, we present an iterative algorithm to solve the second subproblem in \eqref{subproblem}
approximately. Note that this subproblem can be equivalently written as
\begin{equation}\label{sub2}
  \min_\beta \underbrace{\frac{\mu}{2}\left\|X^TX\beta-X^Ty-z^{k+1}+\frac{\lambda^k}{\mu}\right\|_2^2}_{f_k(\beta)}+\|\beta\|_1.
\end{equation}
Since the objective function of \eqref{sub2} is the sum of a smooth
function $f_k$ and the nonsmooth convex function $\ell_1$-norm, the
nonmonotone gradient method II recently proposed by Lu and Zhang
\cite{LuZh09} can be suitably applied to approximately solve
\eqref{sub2}. For ease of reference, we present the algorithm below.
To simplify notations, for any vector $v$ and any real number
$\gamma>0$, we define
\[{\rm
SoftThresh}(v,\gamma):={\rm sgn}(v)\circ \max\left\{0,|v|-\gamma
e\right\}.\]

\noindent {\bf Nonmonotone gradient method:}
\begin{itemize}
  \item[1.] {\bf Start:} Choose parameters $0<\eta,\sigma<1$,
  $0<\underline{\alpha}<1$ and integer $M\ge 0$. Let
  $u^0$ be given and set $\bar\alpha_0=1$.
  \item[2.] {\bf For} $l=0, 1, \ldots$
  \begin{enumerate}[(a)]
    \item Let
    \begin{align*}
      d^l = {\rm SoftThresh}\left(u^l-\bar\alpha_l\nabla f_k(u^l),\bar\alpha_l\right)-u^l, \ \ \
      \
\Delta_l=\nabla f_k(u^l)^Td^l+\|u^l+d^l\|_1-\|u^l\|_1.
    \end{align*}
    \item Find the largest $\alpha\in \{1,\eta,\eta^2,...\}$ such that
    \[f_k(u^l+\alpha d^l)+\|u^l+\alpha d^l\|_1 \le \max_{[l-M]_+\le i\le l}\left\{f_k(u^i)+\|u^i\|_1\right\}+\sigma\alpha\Delta_l.\]
    Set $\alpha_l\leftarrow \alpha$, $u^{l+1}\leftarrow
    u^l+\alpha_ld^l$ and $l\leftarrow l+1$.
    \item Update
    $\bar\alpha_{l+1}=\min\left\{\max\left\{\frac{\|s^l\|^2}{{s^l}^T{g^l}},\underline{\alpha}\right\},1\right\}$,
    where $s^l=u^{l+1}-u^l$ and $g^l=\nabla f_k(u^{l+1})-\nabla
    f_k(u^l)$.
  \end{enumerate}
  {\bf End} (for)
\end{itemize}

\section{Numerical results}\label{sec:numerics}

In this section, we conduct numerical experiments to test the
performance of the ADM for solving problem \eqref{P_orig}. In
particular, we compare our method with the default first-order
method implemented in the TFOCS package \cite{BCG10} for
\eqref{P_orig}.  All codes are written in Matlab and all experiments
are performed in Matlab 7.11.0 (2010b) on a workstation with an
Intel Xeon E5410 CPU (2.33 GHz) and 8GB RAM running Red Hat
Enterprise Linux (kernel 2.6.18).

We initialize the ADM by setting $\beta^0=\lambda^0=0$, and terminate the method once
\begin{equation*}
  \max\left\{\frac{\left|\|\beta^k\|_1-d(\lambda^k)\right|}{\max\{\|\beta^k\|_1,1\}},
  \frac{\|D^{-1}X^T(X\beta^k-b)\|_\infty-\delta}{\max\{\|\beta^k\|_2,1\}},
  \frac{\|X^TX\lambda^k\|_\infty-1}{\max\{\|\lambda^k\|_2,1\}}\right\}\le
  tol
\end{equation*}
for some $tol>0$. For the nonmonotone gradient method subroutine used
to compute $\beta^{k+1}$, we set $\eta=0.5$, $\sigma=1e-4$, $\underline{\alpha}=1e-8$
and $M=1$, and moreover, we initialize the method by setting $u^0=\beta^{k}$. In addition,
we terminate this subroutine once
\begin{equation*}
  \frac{1}{\max\{f_k(u^l)+\|u^l\|_1,1\}}\left\|{\rm SoftThresh}\left(u^l-\nabla
  f_k(u^l),e\right)-u^l\right\|_2 \le  0.1 tol
\end{equation*}
for the same $tol$ as above.% or when the number of (inner) iterations
%reaches $2000$.

For the first-order method implemented in the TFOCS package
\cite{BCG10} for \eqref{P_orig},  we set the restarting parameter to
be $200$ as discussed in \cite[Section~6.1]{BCG10}. We experiment with two
different smoothing parameters: $0.1$ (AT1) and $0.01$ (AT2). We
terminate the first-order method when
\[
\frac{\|\beta^{k+1}-\beta^{k}\|_2}{\max\{\|\beta^{k+1}\|_2,1\}}\le
1e-4.
\]

\subsection{Design matrix with unit column norms}

In this subsection, we consider design matrices with unit column
norms. Similar to \cite[Section~4.1]{CaTa05}, we first generate an
$n\times p$ matrix $X$ with independent Gaussian entries and then
normalize each column to have norm $1$. We then select a support set
$T$ of size $|T|=s$ uniformly at random, and sample a vector $\beta$
on $T$ with i.i.d. entries according to the model $\beta_i =
\xi_i(1+|a_i|)$ for all $i$, where $\xi_i=\pm 1$ with
probability $0.5$ and $a_i\sim N(0,1)$. We finally set
$y=X\beta+\epsilon$ with $\epsilon\sim N(0,\sigma^2I)$.

In our experiment, we choose $\sigma = 0.01$, $0.05$, which
corresponds to $1\%$ and $5\%$ noise, and $(n,p,s) =
(720i,2560i,80i)$ for $i=1,...,10$. For each $(n,p,s)$, we randomly
generate $10$ copies of instances as described above. We then set
$\delta = \sqrt{2\log(p)}\,\sigma$ as suggested by
\cite[Theorem~1.1]{CaTa05}. In addition, we set $\mu =
10/(\sqrt{p}\,\delta)$ and $tol = 1e-3$ for the ADM. Given an
approximate solution $\widetilde \beta$ of \eqref{P_orig}, we
compute a two-stage Dantzig selector $\widehat \beta$ by following
the same procedure as described in \cite[Section~1.6]{CaTa05}, where
we truncate all entries with magnitude below $2\sigma$. We evaluate
the quality of the solutions obtained from different methods by
comparing the following ratios that are introduced in
\cite[Section~4.1]{CaTa05}:
\begin{equation}\label{rho}
  \rho^2_{\rm
  orig}:=\frac{\sum_{j=1}^p(\widetilde
  \beta_j-\beta_j)^2}{\sum_{j=1}^p\min\{\beta_j^2,\sigma^2\}},\ \ \ \rho^2:=\frac{\sum_{j=1}^p(\widehat
  \beta_j-\beta_j)^2}{\sum_{j=1}^p\min\{\beta_j^2,\sigma^2\}}.
\end{equation}
For convenience, we call them the pre-processing and post-processing errors,
respectively. Clearly, the smaller the ratios, the higher the solution quality.
% The number of iterations is reported to the nearest integer, the cpu time is
% reported to the nearest 0.1 seconds, while $\rho^2$ and $\rho^2_{\rm
% orig}$ are reported to the nearest 0.1 digits.

The results of this experiment are reported in Tables~\ref{t1} and \ref{t2}.
In particular, we present the CPU time (cpu), the number of iterations (iter) and
the errors $\rho^2_{\rm orig}$ and $\rho^2$ for all methods, averaged over the $10$
instances. We see from both tables that our ADM generally outperforms the
first-order methods implemented in the TFOCS package \cite{BCG10} in
terms of both CPU time and solution quality. For example, comparing with AT2,
which produces solutions with the best quality among the first-order
methods, our method is about twice as fast and produces solutions
with smaller pre-processing errors and comparable post-processing
errors.

\begin{figure}[t!]
\begin{center}
\caption{\small Recovery result for unit-column-normed $X$ from ADM, $\sigma
= 0.05$}.\label{fig1}
\includegraphics{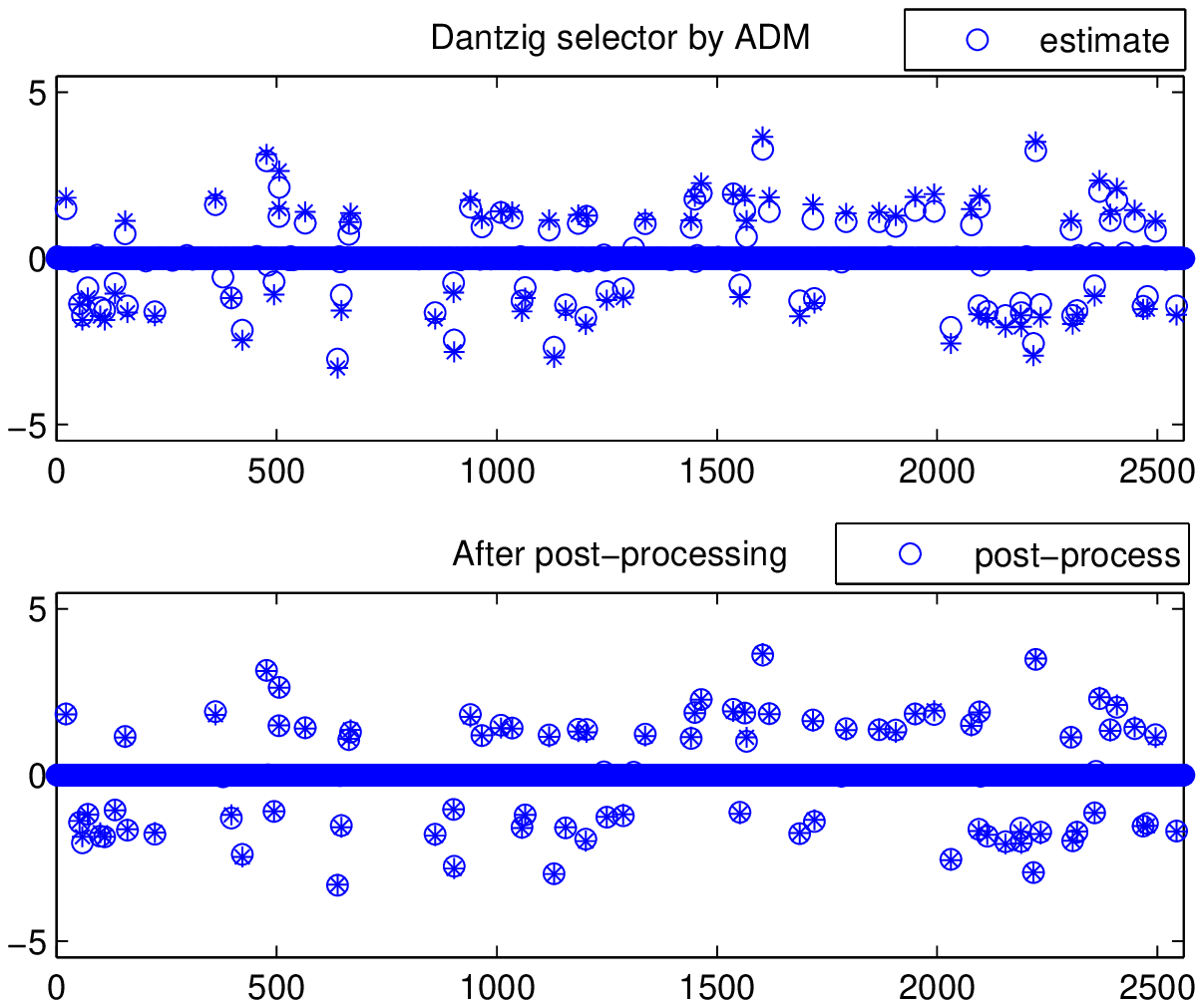}
\end{center}
\end{figure}

\begin{figure}[t!]
\begin{center}
\caption{\small Recovery result for unit-column-normed $X$ from AT2,
$\sigma = 0.05$}.\label{fig2}
\includegraphics{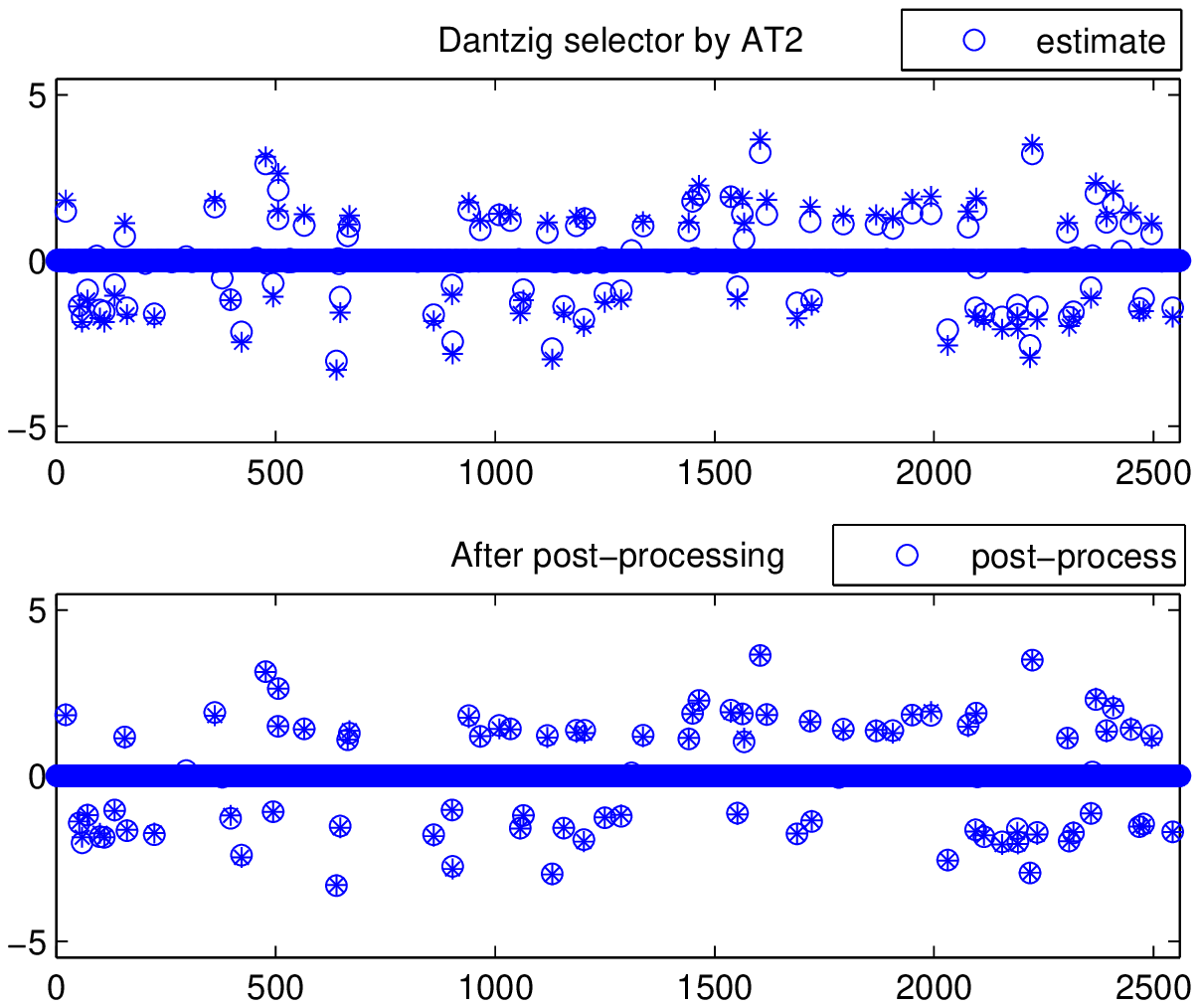}
\end{center}
\end{figure}

In Figure~\ref{fig1}, we present the result for one instance with
size $(n,p,s) = (720,2560,80)$ and $\sigma = 0.05$. The asterisks
are the true values of $\beta$ while the circles are the estimates obtained
by our method before the post-processing (the upper plot) and after
the post-processing (the lower plot). We see from the plot that the
latter estimates are very close to the true values of $\beta$. The similar
phenomenon can also be observed in Figure~\ref{fig2} for the estimates
obtained by AT2 on the same instance.

\begin{table}[t!]
\caption{\small Results for unit-column-normed $X$, $\sigma = 0.01$}
\label{t1}\centering
\begin{footnotesize}
\begin{tabular}{|c c c||r r r||r r r||r r r|}
\hline \multicolumn{3}{|c||}{size} & \multicolumn{3}{c||}{iter} &
\multicolumn{3}{c||}{cpu} & \multicolumn{3}{c|}{$\rho^2$($\rho^2_{\rm orig}$)}\\

\multicolumn{1}{|c}{ $n$} & \multicolumn{1}{c}{ $p$} &
\multicolumn{1}{c||}{ $s$}& \multicolumn{1}{c}{\sc ADM} &
\multicolumn{1}{c}{\sc AT1} & \multicolumn{1}{c||}{\sc AT2}
 & \multicolumn{1}{c}{\sc ADM} &
\multicolumn{1}{c}{\sc AT1} & \multicolumn{1}{c||}{\sc AT2}
 & \multicolumn{1}{c}{\sc ADM} &
\multicolumn{1}{c}{\sc AT1} & \multicolumn{1}{c|}{\sc AT2} \\

\hline

  720  & 2560  &   80  &    13  &   601  &   562   &   2.2  &   7.1  &   6.6   &   1.8(49.2)  &   2.2(88.0)  &   1.9(74.6)\\
 1440  & 5120  &  160  &    12  &   601  &   602   &  11.9  &  32.3  &  32.6   &   1.6(58.2)  &   1.8(82.6)  &   1.4(64.3)\\
 2160  & 7680  &  240  &    27  &   601  &   603   &  35.0  &  67.1  &  67.7   &   1.5(47.7)  &   1.9(83.1)  &   1.5(65.7)\\
 2880  &10240  &  320  &    28  &   601  &   602   &  59.3  & 114.7  & 115.3   &   1.5(52.7)  &   2.0(95.6)  &   1.7(70.8)\\
 3600  &12800  &  400  &    26  &   601  &   643   &  81.2  & 175.3  & 188.2   &   1.5(55.0)  &   1.9(92.2)  &   1.6(69.0)\\
 4320  &15360  &  480  &    28  &   601  &   602   & 119.5  & 250.3  & 251.7   &   1.6(56.2)  &   2.0(95.2)  &   1.6(68.7)\\
 5040  &17920  &  560  &    28  &   601  &   622   & 157.4  & 338.0  & 351.1   &   1.6(58.9)  &   2.0(97.9)  &   1.5(69.8)\\
 5760  &20480  &  640  &    32  &   601  &   622   & 206.9  & 437.7  & 455.0   &   1.5(58.0)  &   1.9(95.4)  &   1.6(70.2)\\
 6480  &23040  &  720  &    37  &   601  &   642   & 268.3  & 552.0  & 592.1   &   1.5(57.6)  &   1.8(93.1)  &   1.5(70.4)\\
 7200  &25600  &  800  &    39  &   601  &   622   & 334.6  & 681.4  & 706.7   &   1.5(57.5)  &   1.9(96.5)  &   1.5(70.3)\\
\hline
\end{tabular}
\end{footnotesize}
\end{table}

\begin{table}[t!]
\caption{\small Results for unit-column-normed $X$, $\sigma = 0.05$}
\label{t2}\centering
\begin{footnotesize}
\begin{tabular}{|c c c||r r r||r r r||r r r|}
\hline \multicolumn{3}{|c||}{size} & \multicolumn{3}{c||}{iter} &
\multicolumn{3}{c||}{cpu} & \multicolumn{3}{c|}{$\rho^2$($\rho^2_{\rm orig}$)}\\

\multicolumn{1}{|c}{ $n$} & \multicolumn{1}{c}{ $p$} &
\multicolumn{1}{c||}{ $s$} & \multicolumn{1}{c}{\sc ADM} &
\multicolumn{1}{c}{\sc AT1} & \multicolumn{1}{c||}{\sc AT2} &
\multicolumn{1}{c}{\sc ADM} & \multicolumn{1}{c}{\sc AT1} &
\multicolumn{1}{c||}{\sc AT2} & \multicolumn{1}{c}{\sc ADM} &
\multicolumn{1}{c}{\sc AT1} & \multicolumn{1}{c|}{\sc AT2} \\

\hline

  720  & 2560  &   80  &    60  &   337  &   601  &   3.8  &   4.0  &   7.2    &   1.4(36.0)  &   1.7(57.1)  &   1.4(37.5)   \\
 1440  & 5120  &  160  &    50  &   340  &   602  &  19.6  &  18.1  &  32.1    &   1.4(43.7)  &   1.7(67.2)  &   1.4(44.3)   \\
 2160  & 7680  &  240  &    39  &   337  &   602  &  33.7  &  37.6  &  67.2    &   1.4(45.7)  &   1.7(67.7)  &   1.4(46.7)   \\
 2880  &10240  &  320  &    49  &   374  &   601  &  64.6  &  71.6  & 115.0    &   1.5(50.7)  &   1.8(75.3)  &   1.4(52.2)   \\
 3600  &12800  &  400  &    52  &   342  &   601  &  96.4  & 100.7  & 177.0    &   1.4(49.9)  &   1.8(75.5)  &   1.4(51.3)   \\
 4320  &15360  &  480  &    56  &   351  &   601  & 131.8  & 146.3  & 251.2    &   1.4(48.9)  &   1.7(72.7)  &   1.4(50.6)   \\
 5040  &17920  &  560  &    57  &   346  &   601  & 170.5  & 194.4  & 337.7    &   1.5(53.1)  &   1.8(78.2)  &   1.4(54.3)   \\
 5760  &20480  &  640  &    60  &   344  &   602  & 207.5  & 251.9  & 440.5    &   1.4(50.9)  &   1.7(73.5)  &   1.4(51.5)   \\
 6480  &23040  &  720  &    60  &   339  &   602  & 251.5  & 312.0  & 554.5    &   1.4(49.9)  &   1.7(74.6)  &   1.4(51.6)   \\
 7200  &25600  &  800  &    64  &   345  &   602  & 309.3  & 390.9  & 683.2    &   1.5(53.1)  &   1.7(79.2)  &   1.4(54.9)   \\
\hline
\end{tabular}
\end{footnotesize}
\end{table}

\subsection{Design matrix with orthogonal rows}

In this subsection, we consider design matrices with orthogonal
rows. We first generate an $n\times p$ matrix $Y$ with independent
Gaussian entries and then set $X$ to be the matrix whose rows form
an orthogonal basis of the row space of $Y$. The vector $y$ is then
generated similarly as in the previous subsection. In particular, we
choose $\sigma = 0.01$, $0.05$, which corresponds to $1\%$ and $5\%$
noise, $(n,p,s) = (720i,2560i,80i)$ for $i=1,...,10$, and $\delta =
\sqrt{2\log(p)}\,\sigma$. For each $(n,p,s)$, we randomly generate
$10$ copies of instances. We set $\mu =1/\delta$ and $tol = 2e-4$
for the ADM.

The computational results averaged over 10 instances are reported in
Tables~\ref{t3} and \ref{t4}. For $\sigma = 0.01$, we observe from
Table~\ref{t3} that our method generally outperforms the first-order
methods implemented in the TFOCS package \cite{BCG10} in terms of
both CPU time and solution quality. In particular, comparing with
AT2, which produces solutions with the best quality among the
first-order methods, our method is at least three times faster and
produces solutions with smaller pre-processing errors and comparable
post-processing errors. On the other hand, for $\sigma = 0.05$, we
see from Table~\ref{t4} that our ADM usually outperforms the
first-order methods in terms of solution quality. In addition, our
method is faster than AT2 which produces solutions with the best
quality among the first-order methods.

\begin{figure}[t!]
\begin{center}
\caption{\small Recovery result for orthogonal $X$, $\sigma =
0.05$}.\label{fig3}
\includegraphics{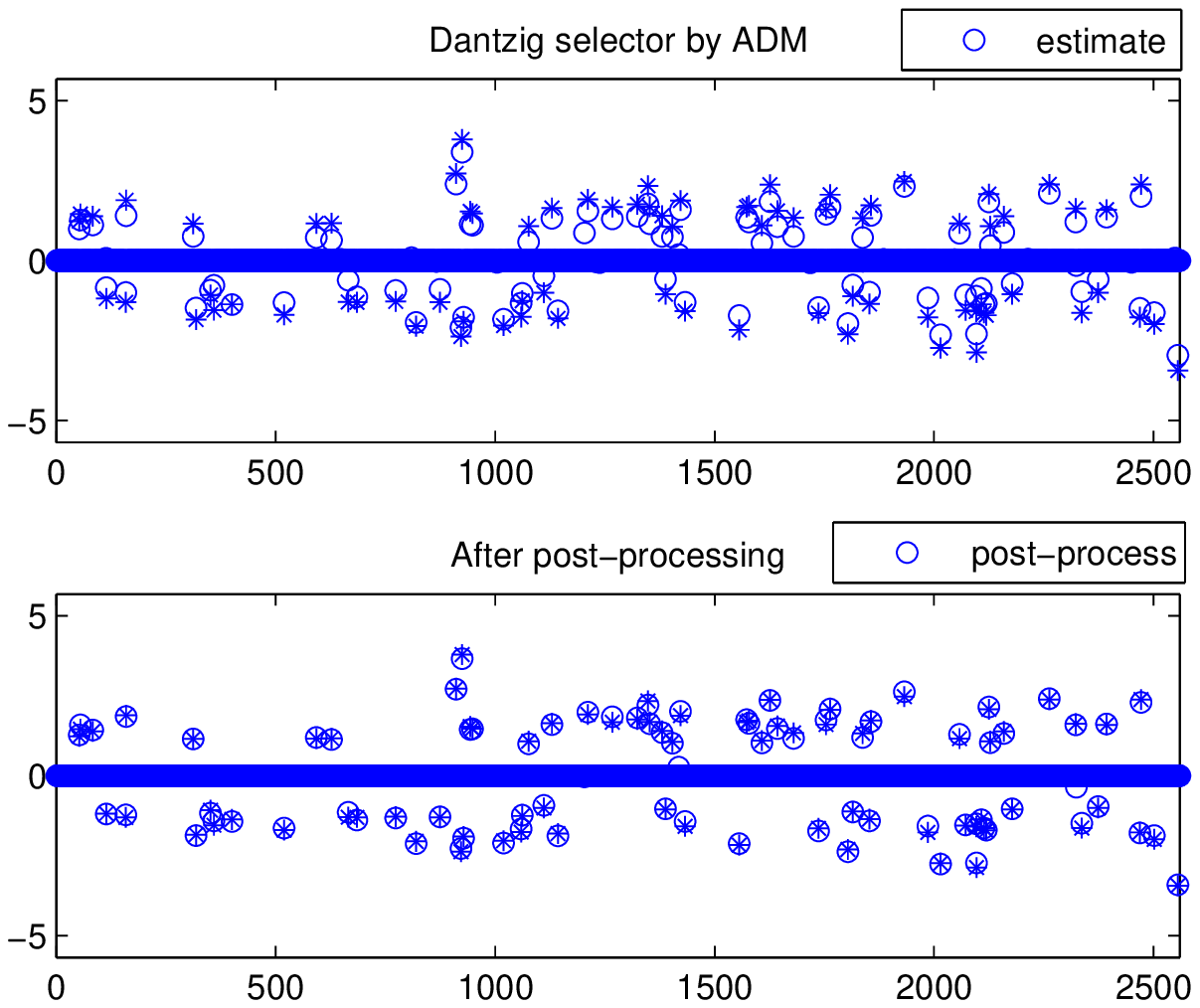}
\end{center}
\end{figure}

\begin{figure}[t!]
\begin{center}
\caption{\small Recovery result for orthogonal $X$, $\sigma =
0.05$}.\label{fig4}
\includegraphics{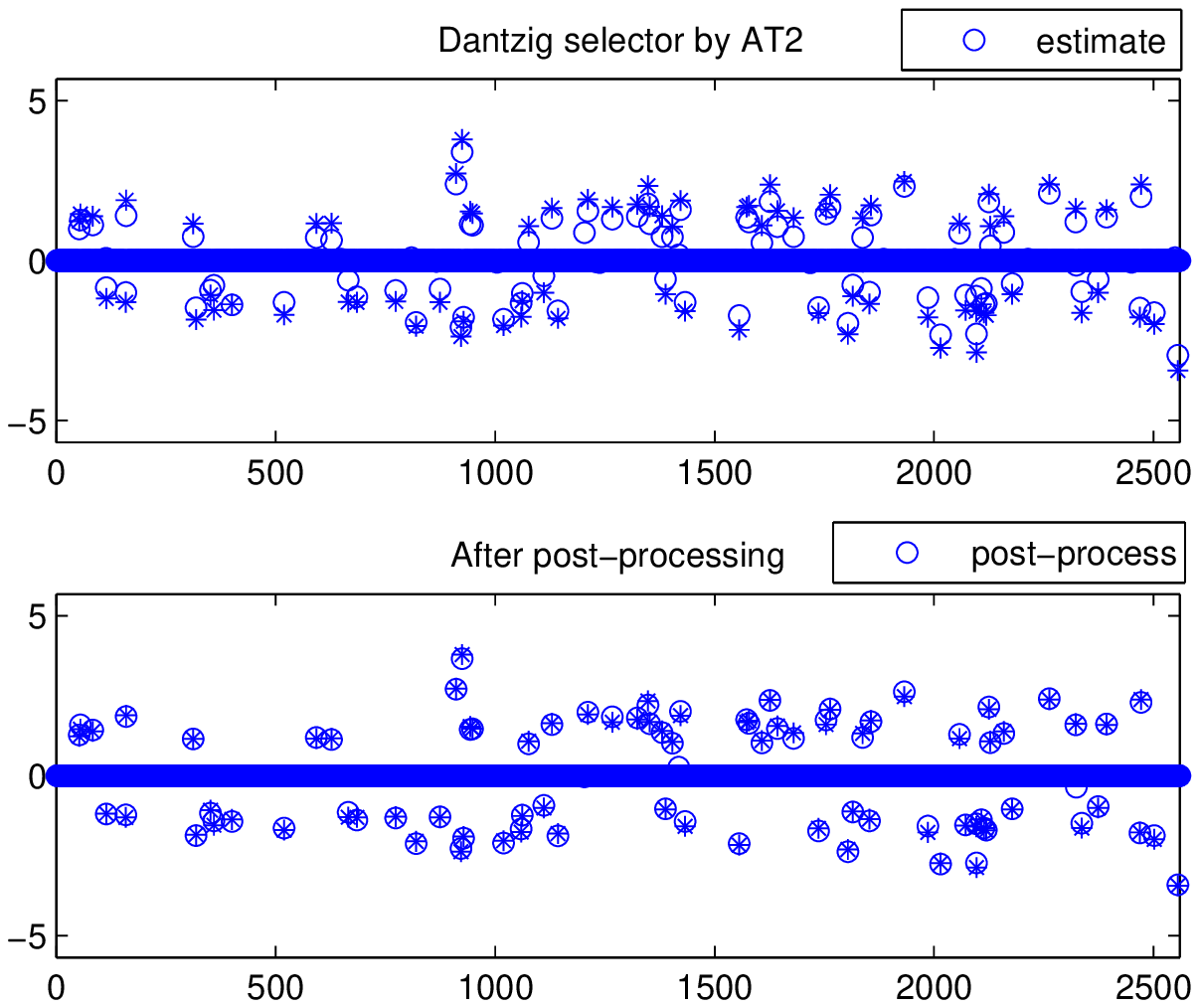}
\end{center}
\end{figure}

In Figure~\ref{fig3}, we present the result for one instance with
size $(n,p,s) = (720,2560,80)$ and $\sigma = 0.05$. The asterisks
are the true values of $\beta$ while the circles are the estimates obtained
by our method before the post-processing (the upper plot) and after
the post-processing (the lower plot). We see from the plot that the
latter estimates are very close to the true values of $\beta$. The similar
phenomenon can also be observed in Figure~\ref{fig4} for the estimates
obtained by AT2 on the same instance.

\begin{table}[t!]
\caption{\small Results for orthogonal $X$, $\sigma = 0.01$}
\label{t3}\centering
\begin{footnotesize}
\begin{tabular}{|c c c||r r r||r r r||r r r|}
\hline \multicolumn{3}{|c||}{} & \multicolumn{3}{c||}{iter} &
\multicolumn{3}{c||}{cpu} & \multicolumn{3}{c|}{$\rho^2$($\rho^2_{\rm orig}$)}\\

\multicolumn{1}{|c}{ $n$} & \multicolumn{1}{c}{ $p$} &
\multicolumn{1}{c||}{ $s$} & \multicolumn{1}{c}{\sc ADM} &
\multicolumn{1}{c}{\sc AT1} & \multicolumn{1}{c||}{\sc AT2} &
\multicolumn{1}{c}{\sc ADM} & \multicolumn{1}{c}{\sc AT1} &
\multicolumn{1}{c||}{\sc AT2} & \multicolumn{1}{c}{\sc ADM} &
\multicolumn{1}{c}{\sc AT1} & \multicolumn{1}{c|}{\sc AT2} \\

\hline

  720  & 2560  &   80  &    32  &   394  &   602   &   1.3  &   4.8  &   7.2  &   5.0(84.2)  &   5.6(105.0)  &   5.1(90.8)   \\
 1440  & 5120  &  160  &    28  &   362  &   602   &   7.5  &  19.4  &  32.3  &   4.3(100.1)  &   5.2(140.7)  &   4.5(111.5)   \\
 2160  & 7680  &  240  &    25  &   388  &   602   &  15.0  &  43.8  &  68.0  &   5.0(102.0)  &   5.8(141.2)  &   4.9(110.5)   \\
 2880  &10240  &  320  &    26  &   398  &   601   &  26.8  &  76.6  & 116.1  &   5.1(113.7)  &   5.8(158.5)  &   5.1(124.8)   \\
 3600  &12800  &  400  &    25  &   389  &   602   &  38.3  & 115.0  & 177.8  &   4.9(110.0)  &   5.5(153.2)  &   4.8(119.9)   \\
 4320  &15360  &  480  &    26  &   388  &   601   &  55.7  & 162.8  & 251.7  &   5.0(110.5)  &   5.6(151.9)  &   4.8(118.8)   \\
 5040  &17920  &  560  &    25  &   389  &   602   &  72.2  & 221.0  & 341.6  &   4.6(112.9)  &   5.3(158.3)  &   4.6(122.0)   \\
 5760  &20480  &  640  &    25  &   384  &   602   &  93.3  & 283.1  & 443.4  &   4.9(114.5)  &   5.5(158.3)  &   4.9(125.8)   \\
 6480  &23040  &  720  &    23  &   394  &   601   & 111.5  & 365.7  & 557.3  &   4.9(117.5)  &   5.5(159.5)  &   4.8(126.5)   \\
 7200  &25600  &  800  &    24  &   392  &   602   & 139.9  & 446.9  & 686.6  &   4.8(116.8)  &   5.5(162.3)  &   4.8(123.6)   \\

\hline
\end{tabular}
\end{footnotesize}
\end{table}

\begin{table}[t!]
\caption{\small Results for orthogonal $X$, $\sigma = 0.05$}
 \label{t4}\centering
\begin{footnotesize}
\begin{tabular}{|c c c||r r r||r r r||r r r|}
\hline \multicolumn{3}{|c||}{size} & \multicolumn{3}{c||}{iter} &
\multicolumn{3}{c||}{cpu} & \multicolumn{3}{c|}{$\rho^2$($\rho^2_{\rm orig}$)}\\

\multicolumn{1}{|c}{ $n$} & \multicolumn{1}{c}{ $p$} &
\multicolumn{1}{c||}{ $s$} & \multicolumn{1}{c}{\sc ADM} &
\multicolumn{1}{c}{\sc AT1} & \multicolumn{1}{c||}{\sc AT2} &
\multicolumn{1}{c}{\sc ADM} & \multicolumn{1}{c}{\sc AT1} &
\multicolumn{1}{c||}{\sc AT2} & \multicolumn{1}{c}{\sc ADM} &
\multicolumn{1}{c}{\sc AT1} & \multicolumn{1}{c|}{\sc AT2} \\

\hline

  720  & 2560  &   80  &   165  &   227  &   440   &   3.3  &   2.7  &   5.1  &   4.9(88.9)  &   5.7(103.2)  &   4.9(92.0)   \\
 1440  & 5120  &  160  &   149  &   225  &   408   &  20.5  &  12.3  &  22.2  &   5.1(97.4)  &   5.9(114.1)  &   5.5(101.3)   \\
 2160  & 7680  &  240  &   137  &   217  &   408   &  40.3  &  24.8  &  46.5  &   5.1(98.4)  &   5.5(112.3)  &   5.1(101.1)   \\
 2880  &10240  &  320  &   129  &   220  &   405   &  65.7  &  42.7  &  78.0  &   4.9(108.4)  &   5.5(125.1)  &   5.0(112.1)   \\
 3600  &12800  &  400  &   145  &   219  &   406   & 110.3  &  65.1  & 120.2  &   5.4(108.3)  &   6.2(126.4)  &   5.7(111.5)   \\
 4320  &15360  &  480  &   132  &   218  &   405   & 140.6  &  92.2  & 170.9  &   5.1(107.4)  &   5.6(121.4)  &   5.3(110.7)   \\
 5040  &17920  &  560  &   125  &   211  &   409   & 182.3  & 121.3  & 233.1  &   4.9(108.6)  &   5.4(124.6)  &   5.1(110.7)   \\
 5760  &20480  &  640  &   115  &   219  &   404   & 219.5  & 162.4  & 298.7  &   4.9(110.4)  &   5.3(126.8)  &   5.0(113.8)   \\
 6480  &23040  &  720  &   133  &   222  &   404   & 310.0  & 207.9  & 376.9  &   5.4(117.3)  &   6.2(134.5)  &   5.5(121.1)   \\
 7200  &25600  &  800  &   119  &   217  &   405   & 345.9  & 249.8  & 464.9  &   5.0(114.8)  &   5.6(130.1)  &   5.1(118.0)   \\

\hline
\end{tabular}
\end{footnotesize}
\end{table}

% \section{Concluding remarks}\label{sec:conclude}

% In this paper, we apply ADM to find Dantzig selectors, and used a
% nonmonotone gradient method proposed in \cite{LuZh09} to
% approximately solve the ADM subproblem. Our computational results
% show that our approach usually outperforms the first-order method
% implemented in the TFOCS package \cite{BCG10} in terms of CPU time
% and solution quality.
% In addition, we observe that our nonmonotone
% gradient method can also be applied to solve the compressed sensing
% problem. It is interesting to compare this approach with some
% existing work in the literature, e.g. \cite{WrNoFi09,WYGZ10}.

\end{document}